# Onion-like Fe$_3$O$_4$/MgO/CoFe$_2$O$_4$ magnetic nanoparticles: new ways to control magnetic coupling between soft/hard phases

*Jorge M. Nuñez* [1,2,3,4], *Simon Hettler*[4,5], *Enio Lima Jr.*[1,2], *Gerardo. F. Goya,*[4,5,6], *Raul Arenal*[4,5,6,7], *Roberto D. Zysler*[1,2,3], *Myriam H. Aguirre*[4,5,6] *and Elin L. Winkler*[1,2,3*]

[1] Resonancias Magnéticas-Centro Atómico Bariloche (CNEA, CONICET) S. C. Bariloche 8400, Río Negro, Argentina

[2] Instituto de Nanociencia y Nanotecnología, CNEA, CONICET, S. C. Bariloche 8400, Río Negro, Argentina.

[3] Instituto Balseiro (UNCuyo, CNEA), Av. Bustillo 9500, S.C. de Bariloche 8400, Río Negro, Argentina.

[4] Instituto de Nanociencias y Materiales de Aragón, CSIC-Universidad de Zaragoza, C/ Mariano Esquillor s/n, Zaragoza 50018, Zaragoza, Spain

[5] Laboratorio de Microscopías Avanzadas, Universidad de Zaragoza, Mariano Esquillor s/n, Zaragoza 50018, Zaragoza, Spain

[6] Dept. Física de la Materia Condensada, Universidad de Zaragoza, C/ Mariano Esquillor s/n, Zaragoza 50018, Zaragoza, Spain

[7] ARAID Foundation, Av. de Ranillas, Zaragoza 50018, Zaragoza, Zaragoza, Spain

**ABSTRACT**

The control of the magnetization inversion dynamics is one of the main challenges driving the design of new nanostructured magnetic materials for magnetoelectronic applications. Nanoparticles with onion-like architecture offer a unique opportunity to expand the possibilities allowing to combine different phases at the nanoscale and also modulate the coupling between magnetic phases by introducing spacers in the same structure. Here we report the fabrication, by a three-step high temperature decomposition method, of $Fe_3O_4/MgO/CoFe_2O_4$ onion-like nanoparticles and their detailed structural analysis, elemental compositional maps and magnetic response. The core/shell/shell nanoparticles present epitaxial growth and cubic shape with overall size of (29±6) nm. These nanoparticles are formed by cubic iron oxide core of (22±4) nm covered by two shells, the inner of magnesium oxide and the outer of cobalt ferrite of ~1 and ~2.5 nm of thickness, respectively. The magnetization measurements show a single reversion magnetization curve and the enhancement of the coercivity field, from $H_C$~608 Oe for the $Fe_3O_4/MgO$ to $H_C$~5890 Oe to the $Fe_3O_4/MgO/CoFe_2O_4$ nanoparticles at T=5 K, ascribed to the coupling between both ferrimagnetic phases with a coupling constant of $J_C$=2 erg/cm$^2$. The system also exhibits exchange bias effect, where the exchange bias field increases up to $H_{EB}$~2850 Oe at 5 K accompanied with the broadening of the magnetization loop of $H_C$~6650 Oe. This exchange bias effect originates from the freezing of the surface spins below the freezing temperature $T_F$=32 K that pinned the magnetic moment of the cobalt ferrite shell.

**INTRODUCTION**

One of the main challenges driving the development of nanostructured magnetic materials is the control of the magnetization response with the magnetic field. The magnetic inversion dynamics, the shape of the hysteresis loop, the coercive field and the saturation magnetization determine the application range of each material which can be tuned for biomedical application, new magnets or magneto-electronic devices.[1] These parameters can be adjusted by combining compounds at the nanoscale with different magnetic characteristic.[2,3] For example, the microfabrication techniques allow the design of exchange coupled magnetic multilayers combining magnetic phases with different magnetic orders as antiferromagnetic (AFM), ferromagnetic (FM) or ferrimagnetic (FiM) and different magnetic anisotropies, that can exhibit both shifting and broadening of the hysteresis loop.[3–7] The coupling between the magnetic phases can be modulated by introducing a non-magnetic spacer that, depending on the thickness, can even decouple them.[8–11] In the latter case, the response of the magnetization with the field presents steps associated with the coercive fields of each phase.[8,12] The combination and manipulation of these features makes it possible to design innumerable devices with different responses for the development of field sensors, MRAM, spin valves, etc.[13–16] Another interesting approach is the design of nanostructures from bottom-up chemical route, which could allow combining in a single nanoparticle phases with different functionalities in an onion-like architecture.[17–19] These architectures permit combining different properties in a single nanoparticle, reducing the size of the functional active unit and also reducing the cost and simplifying the fabrication process.[20] The control of the different synthesis parameters that determine the nanoparticles´ characteristic results in reproducible systems with defined size, interfaces, low dispersion and high crystallinity.[18–21] Different AFM/FiM(FM) or FiM(FM)/AFM core/shell NPs with exchange bias field, coercivity enhancement[22–29], or exchange spring effect[30] have been fabricated and their magnetic properties have been tuned for different applications. There are some reports on the synthesis of systems with enhanced energy product, intended for new rare-earth-free permanent magnets by coating hard magnetic nanoparticles FePt or FePd with a soft magnetic shell Fe, $Fe_3Pt$, $Fe_3O_4$.[20,31–35] Also, core/shell NPs can be applied to magnetic recording devices with smaller elemental magnetic units (bit) and therefore increased density of information, in order to avoid the deleterious effects of thermal fluctuations. This can be achieved by fine-tuning the magnetic anisotropy through a combination of a core/shell (soft/hard) architecture. In this way, the thermal stability can be enhanced and the switching field can be adjusted to invert the magnetization of a bit within the capability of the write head.[20,36] Magnetic NPs can generate building blocks for more complex nanostructures arrangement allowing their integration in thin films for spin-valves design.[37–41] The development of core/shell nanoparticles also finds a wide field of applications in biomedical area. For instance, adjusting the magnetic anisotropy by combining soft and hard magnetic materials in core/shell architecture provides control on the contribution from Brown and Néel relaxation mechanisms to the power absorption for magnetic hyperthermia, optimizing the final heating efficiency.[21,35,42,43] Despite the great potential of design new heterostructures, a few steps have been done to develop more complex magnetic onion-like NPs, in these structures the physicochemical properties are determined by the interfaces and the local characterization at a few nanometer scale is a challenge.[44–48]

In this context and with the aim to move forward the design of novel nanostructures to control the magnetization inversion at the nanoscale, we developed core/shell/shell NPs formed by soft and hard magnetic components separated by a non-magnetic insulator layer. The system was fabricated by a three-step high temperature decomposition method and consists in ~22 nm $Fe_3O_4$ soft magnetic core encapsulated by a MgO intermediate shell of ~1 nm thickness that separates the core from a $CoFe_2O_4$ hard magnetic outer shell of ~2.5 nm thickness. We found that the growing of the third layer results in an enhancement of the coercivity field as a consequence of the coupling of the ferrimagnetic phases even in presence of the MgO separator. The system also exhibits an exchange-bias field which is ascribed to the spin glass order of the $CoFe_2O_4$ surface spins that effectively pin the outer shell spins resulting in an unidirectional exchange anisotropy.

**EXPERIMENTAL SECTION**

**Core/Shell/Shell Nanoparticle Synthesis**

$Fe_3O_4$/MgO/$CoFe_2O_4$ core/shell/shell (CSS) monodispersed nanoparticles (NPs) were synthesized by thermal decomposition of organometallic precursors in presence of oleic acid (OA) and oleylamine (OL) as surfactants in a three-step process based on the method described in [34,49] and illustrated in the Fig.1. Briefly, $Fe_3O_4$ core is synthesized from 3 mmol of iron (III) acetylacetonate (Fe(acac)$_3$) in presence of 1,5 mmol of 1,2-octanediol, 9 mmol of OA, 3 mmol of Ol and 60 mL of 1-octadecene as solvent. Firstly, the solution was heated up to 120 °C during 3 h under a $N_2$ flow to degas the precursors, then the solution was heated up to 200 ºC, and kept there for 10 minutes for core nucleation, and finally it was slowly heated up to promote the growth of the NPs. In a second stage, when the solution reaches 290 ºC, 3 mmol of magnesium acetylacetonate (Mg(acac)$_2$), dissolved in a solution of 1.5 mmol of 1,2-hexadecanediol, 3 mmol of OA, 1 mmol of OL and 15 mL of 1-octadecene, was injected to the solution, to form the first shell, and heat up to 315 ºC for 2 hours. After the cooling process, 30 mL of the resulting solution were separated to overgrow the second shell. Core/shell/shell NPs were prepared in a similar reaction, but in presence of the $Fe_3O_4$/MgO NPs that act as seeds for the growing of the $CoFe_2O_4$ shell. The as-prepared nanoparticles were precipitated by adding 3 times in volume of a solution containing acetone and hexane (14:1) followed by centrifugation (3900 rpm during 30 minutes). Finally, the samples in powder form were dispersed in hexane.

**Structural characterization**

Structural characterization of the systems was performed by means of powder X-ray diffraction experiments in a PANAlytical X´Pert diffractometer with Cu Kα radiation using a glass sample holder. The crystalline structure, morphology and size dispersion of the NPs were analyzed by transmission electron microscopy (TEM) and high-resolution TEM (HRTEM) with an aberration corrected Titan$^3$ 60–300 (ThermoFisher Scientific, formerly FEI) microscope operating at 300 kV at room temperature. High-resolution scanning TEM (HRSTEM) images acquired with a high-angle annular dark field (HAADF) detector (Fischione) were obtained in a $C_S$-probe-corrected Titan (ThermoFisher Scientific, formerly FEI) at a working voltage of 300 kV. Electron energy-loss spectra (EELS) were acquired in this Cs-probe corrected microscope using a Tridiem Energy Filter (Gatan) spectrometer at an energy dispersion of 0.5 eV/pixel. Spectrum images were acquired with 500 ms dwell time and a pixel step size of 0.7 nm. The collection semiangle (β) was 51.3 mrad for a camera length of 10 mm and a spectrometer entrance aperture of 1 mm. The convergence semiangle (α) was 24.8 mrad. The energy resolution, estimated from the full width at half-maximum (FWHM) of the zero-loss peak, was 0.8 eV.

**Magnetic characterization**

The magnetic properties were studied by means of a commercial superconducting quantum interference device magnetometer (SQUID, MPMS Quantum Design). The magnetization was measured as function of temperature using the field-cooling (FC) and zero-field-cooling (ZFC) protocols, with a field from 50 Oe to 50 kOe. Magnetization as function of an applied field up to 50 kOe was measured with a ZFC and FC protocols; in this last case the samples were cooled from room temperature down to the measured temperature with an applied field of 10 kOe. To perform the magnetic measurements, 3 mg of nanoparticles were dispersed and fixed in 1 g of epoxy resin to reduce the interparticle interactions and to suppress mechanical movement of the NPs. In order to normalize the magnetization with the magnetic nanoparticles mass, the proportion of the organic compound in the as-made nanoparticles was determined by means of thermogravimetric analysis (TGA) in a Shimadzu DTG-60H equipment. AC susceptibility measurements were performed in a Quantum Design PPMS ac/dc magnetometer using an excitation field of Hac = 4 Oe and frequencies $1 \text{ Hz} \leq f \leq 1.5$ kHz, as a function of temperature.

**RESULTS AND DISCUSSION**

Figure 1 schematizes the three-step seed-mediated high temperature decomposition synthesis route. This figure indicates the first synthesis stage where the nucleation and growth of the $Fe_3O_4$ core take place. In the second stage the respective precursors, which are detailed in the experimental section, are hot injected in order to nucleate and grow the MgO over the cores, and then the solution was cooled down. Finally, in the third stage, the solution is heated up and the precursors to grow the second shell of $CoFe_2O_4$ were hot injected as signaled in the figure. It is important to remark that proper solvents should be selected to reach the reflux condition at a higher temperature than the decomposition temperature of the organic precursors. This guarantees to reach the metal ions supersaturation condition and the nucleation of the metal oxides phase. The $Mg(acac)_2$ precursor has a decomposition temperature of T~538 K, larger than the $Fe(acac)_3$ (T~453 K) and $Co(acac)_2$ (T~440 K), determining the use of 1-octadecene as solvent (T~587 K). The requested reflux condition also restricts the use of surfactants to those with larger decomposition temperature, such as 1,2-hexadecanediol (T~ 576 K). Representative low and HRTEM images of the $Fe_3O_4$/MgO sample obtained in the second stage of the synthesis, are shown in Fig. 2-(a) and (b). From this figure, nanoparticles with cubic shape and uniform size can be observed. The size histogram shown in Fig. 2-(c) was obtained by measuring more than 300 NPs. The mean diameter and size dispersion was obtained from the fitting with a log-normal distribution $f(d) = (\sigma d\sqrt{2\pi})^{-1} e^{-[ln^2(d/d_0)/2\sigma^2]}$, from where the mean diameter $<d> = d_0 \, e^{\sigma^2/2}$ and the standard deviation $\sigma = <d> \left[e^{\sigma^2-1}\right]^{1/2}$ were calculated, resulting (24±4) nm for the core/shell NPs. Figure 2-(d) and (e) shows the $Fe_3O_4/MgO/CoFe_2O_4$ nanoparticles obtained after the third stage of the syntheses, where again cubic shape nanoparticles are observed. Figure 2-(f) shows the size histogram, constructed by measuring the size of more than 300 NPs, fitted with a log-normal distribution from where the mean nanoparticle size and size dispersion was obtained (29±6) nm. The analysis from high resolution TEM (HRTEM) images reveals that the nanoparticles are single crystalline, and the successive layers grew epitaxially due to the negligible lattice mismatch between the $Fe_3O_4$, MgO and $CoFe_2O_4$ (≈0.34-0.38%). The analysis of the images also showed interplanar distances d=2.10(2) Å and d=2.94(2) Å that could be identified in the core/shell and core/shell/shell nanoparticles, consistent with the interplanar distance of the (400) and (220) planes of the spinel phase, respectively. Also, the distance d=2.11(2) Å agrees well with the corresponding (200) crystalline plane of the MgO phase. The fast Fourier transformation (FFT) images of the core/shell and core/shell/shell confirmed the single-crystal growth with the plane indexation shown in the inset of Fig.2-(b) and (d). These features are also evidenced in the XRD pattern of the $Fe_3O_4/MgO/CoFe_2O_4$ nanoparticles, Fig. 3, where the overlap of the peaks corresponding to the iron and cobalt spinel and the MgO phase are observed, with no other detected phases. Using the diffraction peaks position, the spinel lattice parameter, $a_{spinel}$, was calculated from the relationship between the Miller indices *(h,k,l)* and the corresponding interplanar distance $d_{hkl}$ for a cubic structure: $a=d_{hkl}(h^2+k^2+l^2)^{1/2}$, resulting *a*=0.841(3) nm.

Analogously, the lattice parameter of the magnesium oxide $a_{MgO}$=0.422(2) nm was obtained. The calculated lattice parameters, $a_{spinel}$ and $a_{MgO}$, are in agreement with the reported for magnetite ($a_{Fe3O4}$=0.8392 nm), cobalt ferrite ($a_{CoFe2O4}$=0.8392 nm) and magnesium oxide ($a_{MgO}$=0.4211 nm) [5,50]. Also the crystallite size of the core/shell/shell NPs was obtained from the x-ray powder pattern. To perform this analysis the peaks were fitted with a pseudo-Voigt function in order to obtain the full width at half maximum (FWHM), and then the crystallite size was calculated by using the Scherrer equation, resulting in a median value of 22(2) nm. The smaller crystallite size obtained from XRD compared to the mean size obtained by TEM indicates the presence of surface disorder in the core/shell/shell structure.

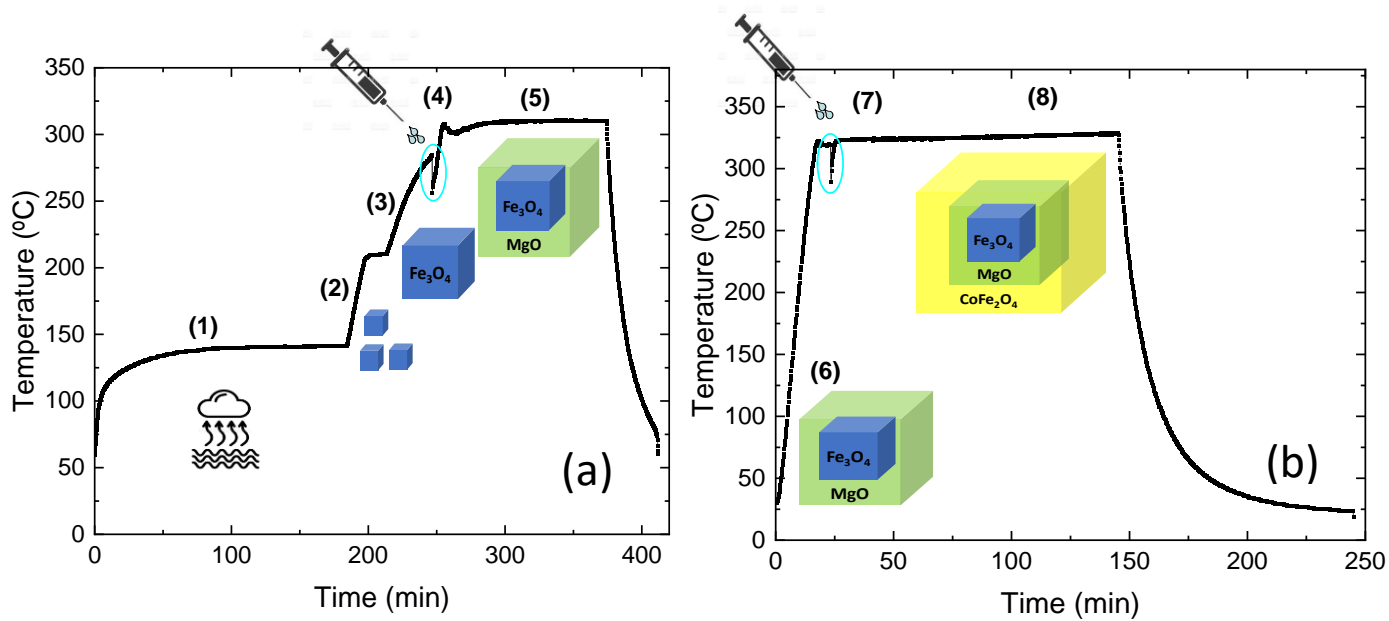

**Figure 1:** Temperature ramp profile used in the three-step high temperature decomposition method and schematic illustration of the Fe$_3$O$_4$/MgO/CoFe$_2$O$_4$ nanoparticles growth. Left panel: In the first synthesis step the precursors were heating (Fe(acac)$_3$, 1,2-octanediol, OA, Ol and 1-octadecene) up to 120 °C for 3 h under N$_2$ flow (1), then the solution was heated up to 200 °C for 10 minutes for the Fe$_3$O$_4$ cores nucleation (2), followed by a ramp to 290 °C to promote the growth of the Fe$_3$O$_4$ NPs (3). In the second stage Mg(acac)$_2$, 1,2-hexadecanediol,

OA, OL and 1-octadecene were injected at 290 ºC (4), then the solution was heated up to 315 ºC for 2 hours to grow the MgO shell (5), followed by the cooling process. Right Panel: Sketch of the final synthesis step including the heating of the solution containing $Fe_3O_4$/MgO NPs seeds up to 315 ºC (6), where the precursors $Fe(acac)_3$, $Co(acac)_2$, OA, OL and 1-octadecene were injected (7) for the growing of the $CoFe_2O_4$ outer shell (8).

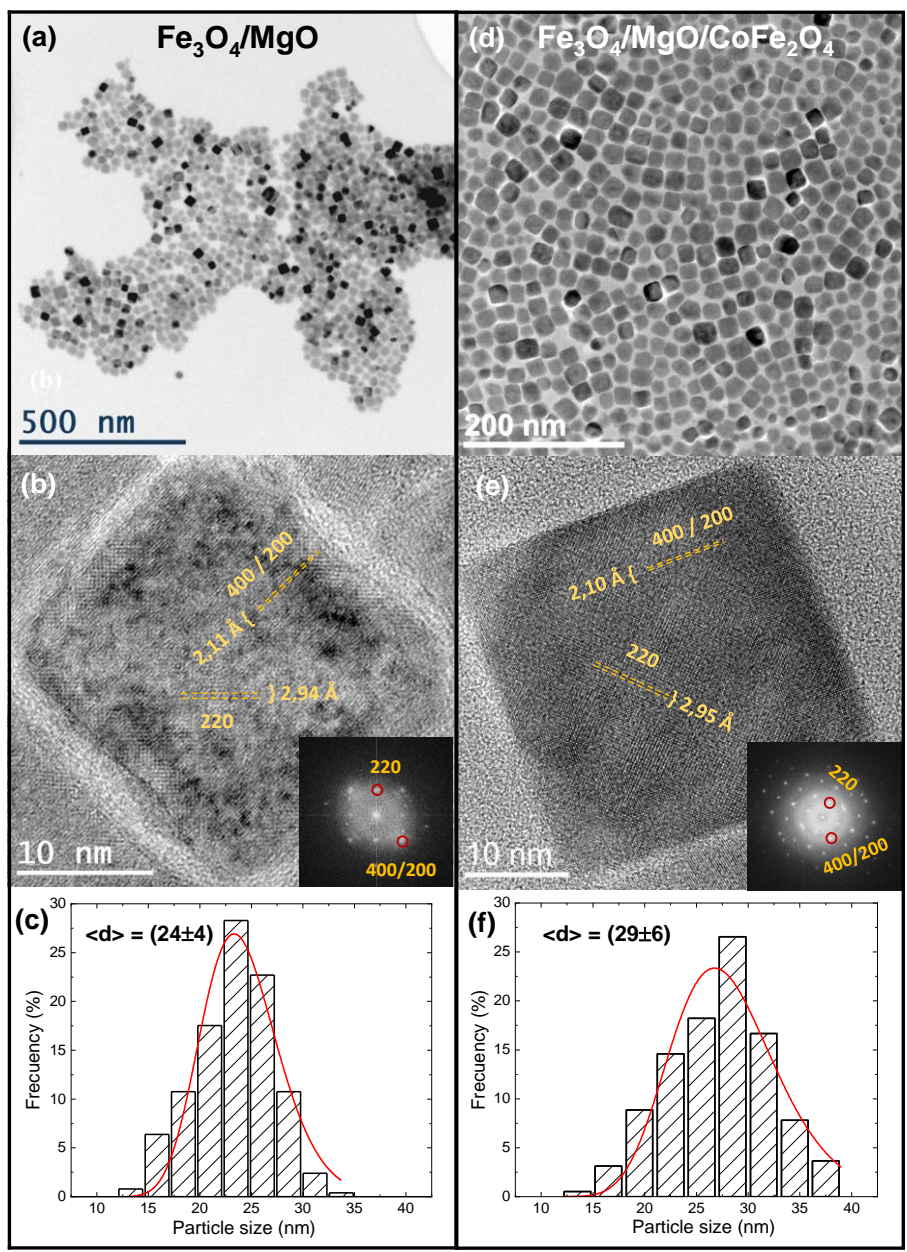

**Figure 2:** Bright-field TEM micrographs of (a) Fe3O4/MgO and (d) Fe3O4/MgO/CoFe2O4 nanoparticles. Aberration-corrected HRTEM image of a (b) Fe3O4/MgO nanoparticle and (e)

Fe3O4/MgO/CoFe2O4 nanoparticle and their corresponding FFT images of the whole NPs are shown in the insets. Size histograms fitted with a log-normal distribution for (c) $Fe_3O_4$/MgO and (f) $Fe_3O_4$/MgO/$CoFe_2O_4$ nanoparticles. From the fitting (24±4) nm and (29±6) nm were obtained for the core/shell and core/shell/shell, respectively, corresponding to an CoFe2O4 outer layer of 2.5 nm thickness.

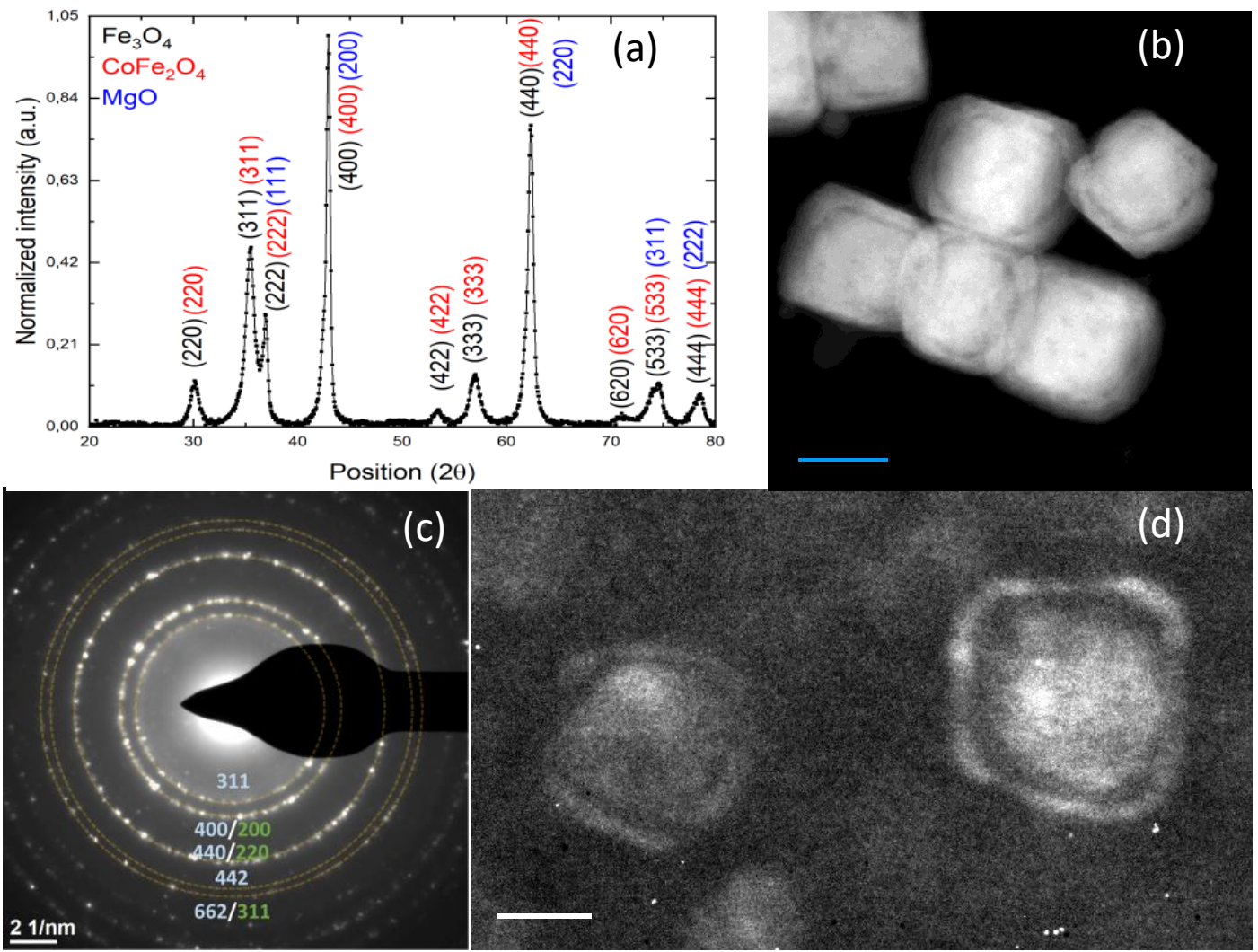

**Figure 3:** (a) XRD pattern of $Fe_3O_4$/MgO/$CoFe_2O_4$ nanoparticles, where the diffraction peaks of the $Fe_3O_4$ (black), $CoFe_2O_4$ (red) and MgO (blue) are indexed. (b) STEM-HAADF image where the annular magnesium oxide shell can be detected by Z contrast, showing core of size ~22 nm coated with an intermediate (MgO) thin layer of ~1 nm thickness and an outer

thicker layer (CoFe$_2$O$_4$) of ~2.5 nm thickness. (c) Selected area diffraction pattern, where the rings were indexed with the Fm3m MgO (green) and Fd3m Fe$_3$O$_4$/ CoFe$_2$O$_4$ (cyan) space groups. (d) Dark-field TEM images of Fe$_3$O$_4$/MgO/CoFe$_2$O$_4$ NPs selecting a section of the (311) spinel diffraction ring using a small objective aperture.

Due to the epitaxial growth and the negligible mismatch between the different phases, the onion-like architecture could not be resolved from HRTEM lattice-fringe images (Figure 2-(b) and (d)). However, dark-field images, selecting a section of the (311) spinel diffraction ring using a small objective aperture, showed a bright contrast only for the Fe$_3$O$_4$ and CoFe$_2$O$_4$ spinel phases, unveiling the core/shell/shell structure as shown in Fig. 3-(c). This onion-like architecture was confirmed by HAADF-STEM imaging that is a known technique for material characterization with high spatial resolution and with a contrast proportional to ~$Z^{1.7}$, known as Z-contrast.[51–53] Figure 3-(b) shows representative HAADF-STEM images where a clear dark annular contrast is observed in the inner shell corresponding to the MgO phase. From these measurements a mean core nanoparticle size (22±4) nm and the thickness of the inner MgO shell of ~1 nm was measured. It is worth to mention that from the comparison between the size histograms of core/shell and core/shell/shell (see Fig. 2-(c) and (f)) the thickness of the outer shell could be calculated, obtaining a value of ~2.5 nm.

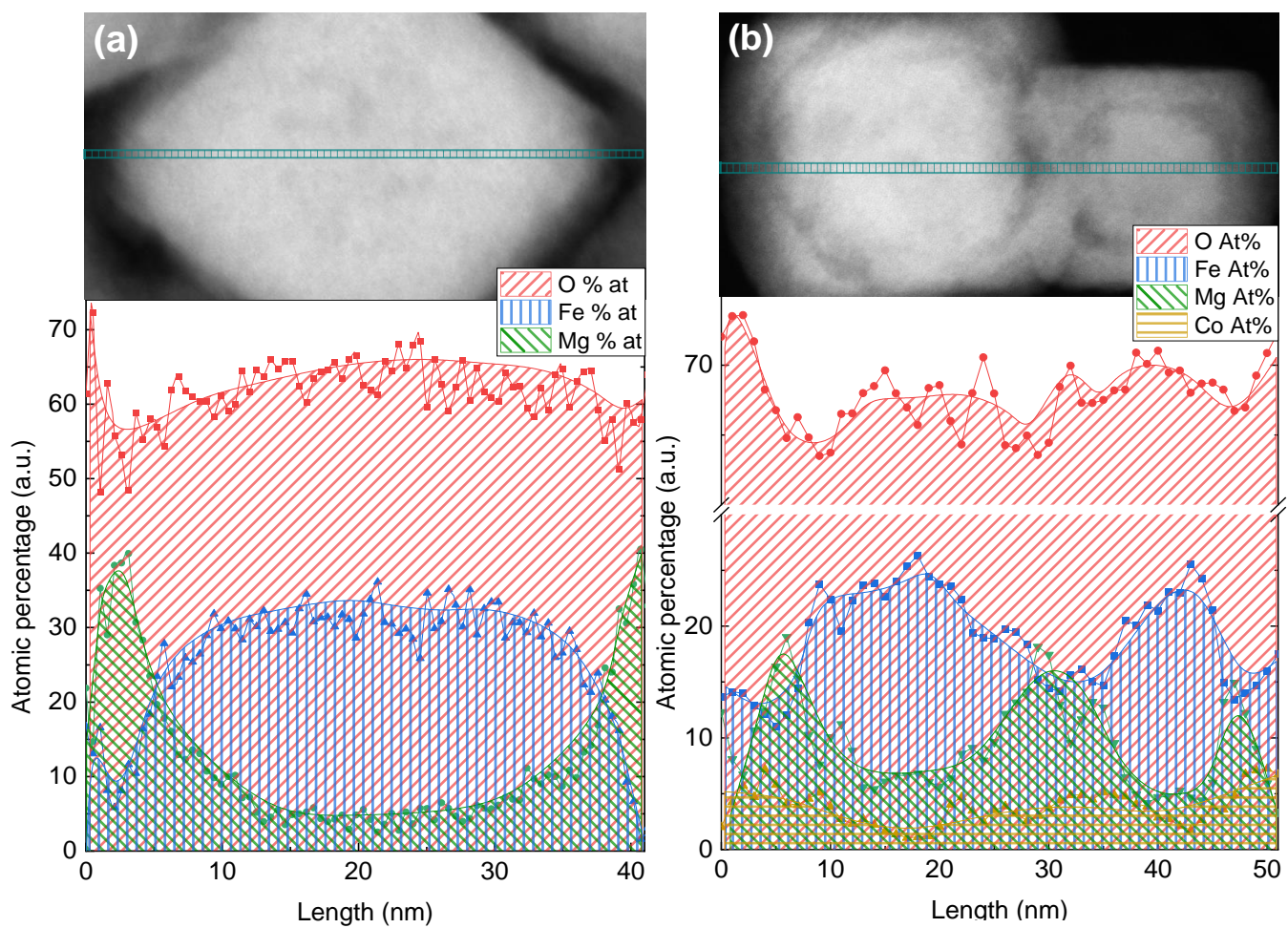

**Figure 4:** Elemental profile performed by EDS on (a) isolated $Fe_3O_4/MgO$ nanoparticles and (b) $Fe_3O_4/MgO/CoFe_2O_4$ nanoparticles (O in red + Fe in blue + Mg in green + Co in yellow).

Elemental mapping of the core/shell and core/shell/shell nanoparticles were analyzed by energy dispersive X-ray spectroscopy (EDS) and electron energy-loss spectroscopy (EELS). Figure 4-(a) shows an EDS elemental profile on a single core/shell nanoparticle, where Fe, Mg and O were detected. From the atomic percentage elemental profile, the $Fe_3O_4/MgO$ structure is corroborated. Analogously, Fig.4-(b) shows the EDS profile scanning over two nanoparticles confirming the onion-like architecture. More accurate nanoparticles elemental mapping was investigated by EELS and a representative spectrum is shown in Fig. 5-(e). In this figure the peaks associated to O-K edge (red), Fe-L edge (blue), Co-L edge (yellow) and Mg-K edge (green) are identified in the spectrum. By performing spectrum imaging (SI) and integration of the background-subtracted edge areas in every pixel, the spatial elemental distribution was obtained (Fig.5-(a-d)). The maps confirm the highest concentration of magnesium in the inner shell and cobalt in the outer shell consistently with the core/shell/shell architecture. This observation can be quantified by a linear profile composition analysis, shown in Fig. 5-(g), confirming the increase of magnesium with the corresponding decline of iron oxide in the intermediate shell, and also the increase of cobalt and iron in the outer shell. Notice that, in some interface sections, a thinning of the MgO intermediate layer can be observed, Fig. 5-(e). The lineal profile atomic distribution, Fig.5-(g), also shows that the magnesium is concentrated in a ring with average diameter of ~23 nm and its concentration extends to the nanoparticle edge, suggesting that the cobalt ferrite oxide is doped with magnesium. The doping of the outer shell with magnesium could be related to either interface interdiffusion during the synthesis, or the presence of Mg excess ions remaining from the second stage of the synthesis. The core/shell/shell architecture of the NPs is also revealed by an energy loss near edge structure (ELNES) analysis of the O-K core-loss edge shown in Fig. S1 of the Supporting Information. This figure presents the comparison of the O-K edge at 530 eV obtained from core, core-shell and core-shell-shell areas of a NP. The three spectra are compared with a $Fe_3O_4$ magnetite and a MgO reference.[47,54–57] The major difference is observed in the pre-peak located at 530 eV, which is strong in the core region and the outer shell, while shows a decreased intensity in the inner MgO shell. This observation corresponds well to the reference spectra, as magnetite shows a prominent pre-peak, while it is completely absent for MgO. The spectrum of the outer shell does not perfectly agree with the magnetite reference, as the pre-peak intensity is reduced in cobalt ferrite when compared to $Fe_3O_4$,[47] and also due to the presence of Mg that it is expected to alter the O-K edge. These results have important implications in the magnetic response of the nanoparticles discussed below.

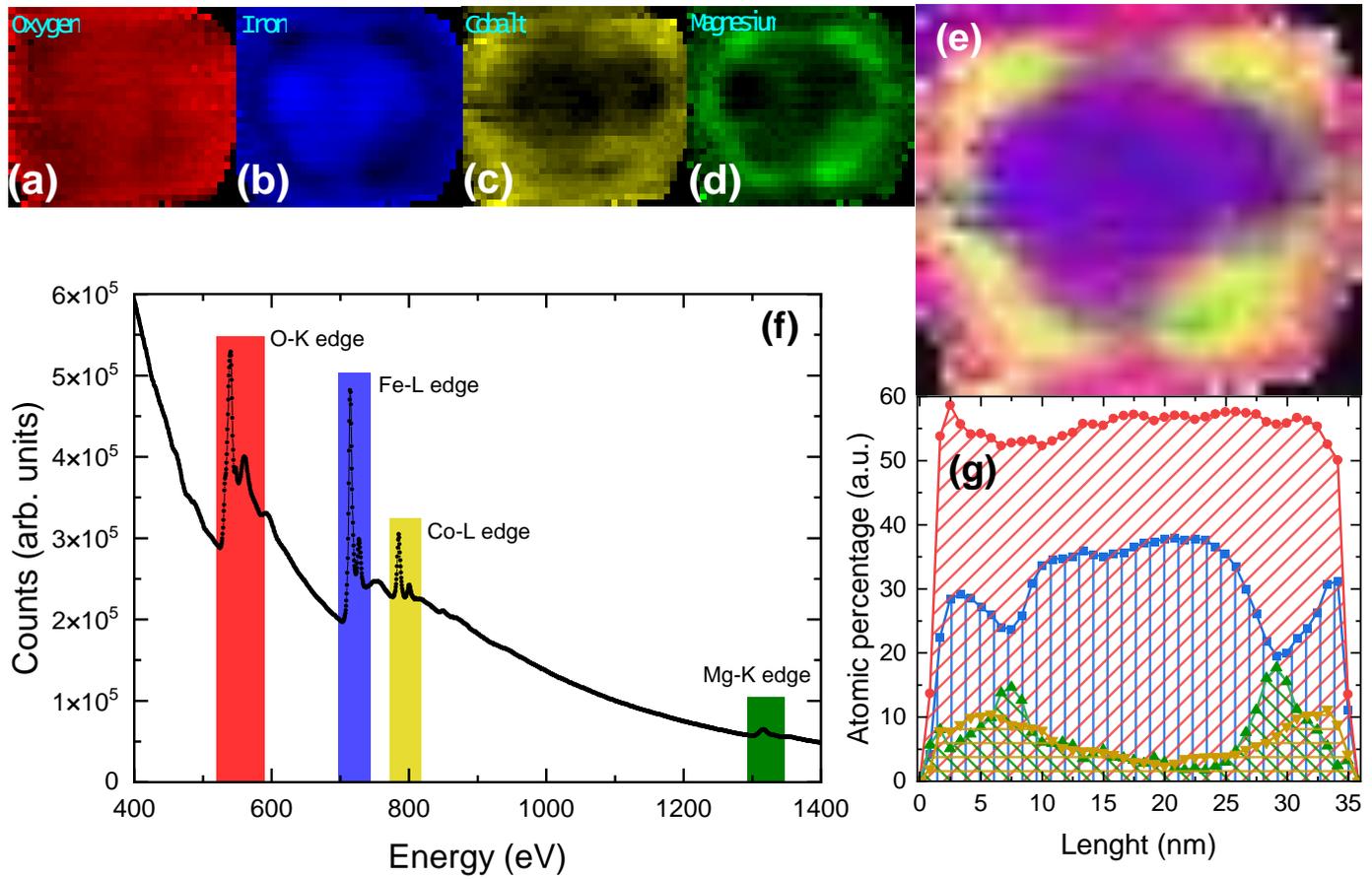

**Figure 5.** Elemental mapping performed by EELS-SI on an isolated $Fe_3O_4$/MgO/$CoFe_2O_4$ NP of (a) O in red, (b) Fe in blue, (c) Co in yellow, (d) Mg in green and (e) the composition map. (f) EELS sum spectra of SI with the O−K edge (532 eV, red vertical line), Fe−L (713 eV, blue vertical line) edges, Co−L edge (781 eV, yellow vertical line) as well Mg−K edge (1323 eV, green vertical line) are indicated. (g) Line profile across the NP, extracted from this EELS dataset, of atomic percentages, where the core/shell/shell structure is evidenced.

Figure 6 compares the field dependence of the magnetization at 5K of the $Fe_3O_4$/MgO and $Fe_3O_4$/MgO/$CoFe_2O_4$ nanoparticles. From this measurement the enhancement of the coercivity field is clearly observed from $H_C$=608 Oe for the core/shell NPs to $H_C$=5890 Oe when the third shell of cobalt ferrite is grown. Figure 6-(b) shows the magnetization loops as a function of the temperature in the ZFC condition for the $Fe_3O_4/MgO/CoFe_2O_4$ nanoparticles. It is noteworthy that a single reversion curve is observed for all the temperatures, signaling that the FiM phases are coupled even though they are separated by a MgO diamagnetic insulator interlayer. This result is consistent with the magnetic response reported by Zaag et al. for multilayers.[8,9] These authors studied the coupling in $Fe_3O_4/MgO/Fe_3O_4/Co_xFe_{3-x}O_4$ thin films as a function of the MgO thickness ($t_{MgO}$) and identified two different coupling regime: i) a weaker interlayer interaction for $t_{MgO}$ >1.3 nm, where a stepped hysteresis loop is observed due to the different coercive fields of the magnetic layers and ii) a rapid increase of the coercivity for $t_{MgO}$ <1.3 nm due to the enhancement of the ferromagnetic coupling between the layers, where the magnetization loop tends to a single reversion magnetization behaviour.[8,9] In this work, the authors assumed that the change from weak to strong coupling is due to irregularities at the interface, where $Fe_3O_4$ bridges through the MgO are formed. In the present case, the magnetic reversion curve of the $Fe_3O_4/MgO/CoFe_2O_4$ nanoparticles ($t_{MgO}$~1 nm) is consistent with the behaviour observed in multilayers for the strong coupling regime ($t_{Mg}$<1.3 nm). Moreover, Fig. 5 shows the presence of ferrimagnetic bridges through the MgO shell, confirming the hypothesis presented to explain the increasing coupling in nanostructures for thinner spacers. The magnitude of the surface coupling energy ($J_C$) can be estimated from the difference in the coercivity fields between the uncoupled $Fe_3O_4$ NPs ($H_C^{core}$) and the coupled onion ($H_C^{onion}$) system:[9,58]

$$J_C = (H_C^{onion} - H_C^{core})t^{core}M_S^{core}, \quad (1)$$

where $t^{core}$ and $M_S^{core}$ are the thickness and saturation magnetization of the non-interacting Fe$_3$O$_4$ phase. From this equation $J_C$= 2 erg/cm$^2$ was calculated using $H_C^{core} = 608\,Oe$, $H_C^{onion} = 5890\,Oe$, $t^{core}$=22 nm, $M_S^{core}$=35 emu/g. This value is larger than the obtained for the Fe$_3$O$_4$/MgO/Fe$_3$O$_4$/Co$_x$Fe$_{3-x}$O$_4$ multilayers in the strongly-coupled regimen, $J_C$~0.3 erg/cm$^2$,[8] probably because the surface-to-volume ratio of the Fe$_3$O$_4$ core is 10 time larger than the Fe$_3$O$_4$ phase in the multilayer, resulting in a larger effective coupling surface; furthermore heterogeneous seed mediated growth of core/shell/shell architecture could results in larger interface imperfections, in particular in the thinner MgO shell, than in the multilayer fabricated by molecular beam deposition. For uniaxial, randomly oriented and non-interacting nanoparticles where the magnetization reverts coherently with the magnetic field it is predicted that the coercive field follow the relation $H_C(T) = H_C^0 \left[1 - \left(\frac{T}{T_B}\right)^{1/2}\right]$, where $H_C^0$ is the coercive field extrapolated at zero temperature value.[59] Despite the complexity of this onion-like nanoparticles system, Fig. 6-(c) shows that the coercive field follows a T$^{1/2}$ dependence supporting the single magnetization reversion of the NPs due to the strong coupling of the ferrimagnetic phases. From the fitting curve $H_C^0$=(7033±264) Oe and $T_B$=(216±37) K were obtained. On the other hand, the calculation of the effective magnetic anisotropy from the blocking temperature ($T_B = 25 K_{eff} V/k_B$); in this three-layer NPs is not straightforward due to the presence of a non-magnetic interlayer.

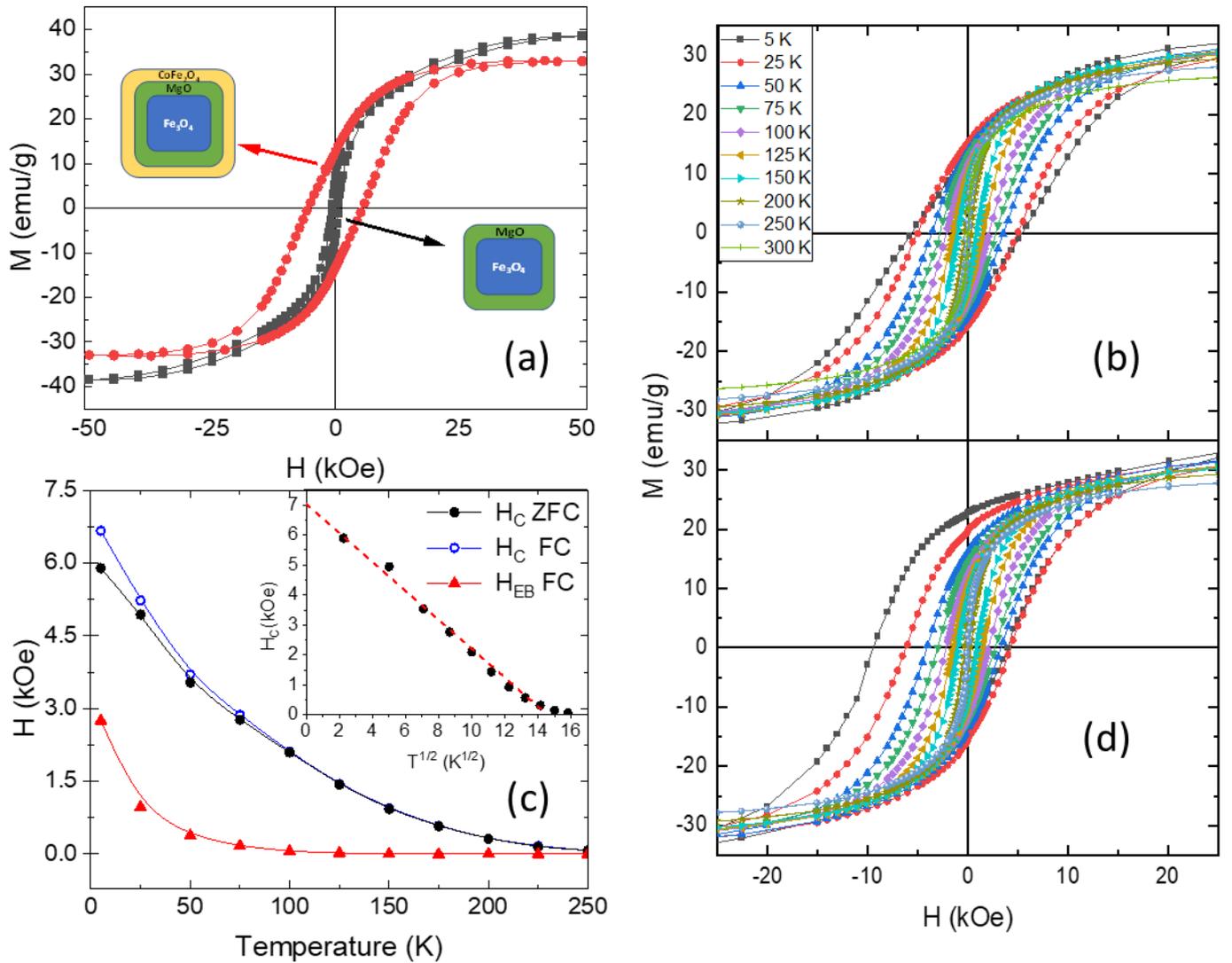

**Figure 6.** (a) Hysteresis cycles of the $Fe_3O_4$/MgO and $Fe_3O_4$/MgO/$CoFe_2O_4$ NPs systems, measured at 5 K. (b) ZFC and (d) FC, from 320 K with 10 kOe, hysteresis loops of $Fe_3O_4$/MgO/$CoFe_2O_4$ NPs systems measured in the 5 K – 300 K temperature range. (c) Temperature dependence of the $H_C$ measured with the ZFC (black dot) and FC (blue dot) protocols, and $H_{EB}$ (red triangle) of the $Fe_3O_4$/MgO/$CoFe_2O_4$ system. The inset shows the dependence of $H_C(T)$ with $T^{1/2}$.

Figure 6-(d) shows the magnetization loops measured after cooling the $Fe_3O_4/MgO/CoFe_2O_4$ sample from room temperature without a magnetic field (ZFC) and with an applied field of 10 kOe (FC). From this figure a single reversion curve is observed in the whole temperature range and also a systematic increase of the coercive field as the temperature decreases. Moreover, in the FC magnetization measurements a clear shift toward negative field is observed for temperature T<75 K evidencing the presence of exchange bias effect. It is important to remark that no exchange bias effect was observed in the $Fe_3O_4/MgO$ core/shell system. Figure 6-(c) shows the temperature evolution of the exchange bias field which grows up to 2850 Oe at 5 K. The shifting of the hysteresis cycles is also accompanied by an enhancement of the coercivity field. It is known that the exchange bias effect is present in nanoparticles with AFM/FM (FIM) interfaces[3,26,60,61] and also in systems with interface exchange coupling between the magnetically ordered core with disordered and frozen surface spins.[62–65] In these systems the FM (FiM) phase has pinned spins at the interfaces due to the coupling with the more anisotropic AFM state or with the surface spin glass phase. Even larger exchange bias and coercivity field enhancement were found when the ferromagnetic phase is coupled with the more disordered spin glass state, when compared with the coupling with AFM phase, indicating a larger amount of pinned spins at the interface.[66] Based on the compositional and the morphological analysis of the $Fe_3O_4/MgO/CoFe_2O_4$ nanoparticles, the origin of the exchange bias is ascribed to the formation of spin glass-like states at the outer FiM shell. This hypothesis is supported by previous reports, in particular the magnetic behaviour of ferromagnetic hollow nanoparticles whose morphology is similar to the $CoFe_2O_4$ shell growth over the non-magnetic MgO shell.[67–69] These systems present a larger degree of spins surface disorder that freeze at low temperature increasing its surface anisotropy and showing large exchange bias effects as a consequence of the magnetic coupling with the FiM order phase. In addition, the doping of the $CoFe_2O_4$ shell with non-

magnetic magnesium ions, as shown by the EELS analysis, introduce a larger degree of magnetic disorder spins that froze at lower temperature. In order to support this picture, we analyze the evolution of the magnetization with the temperature.

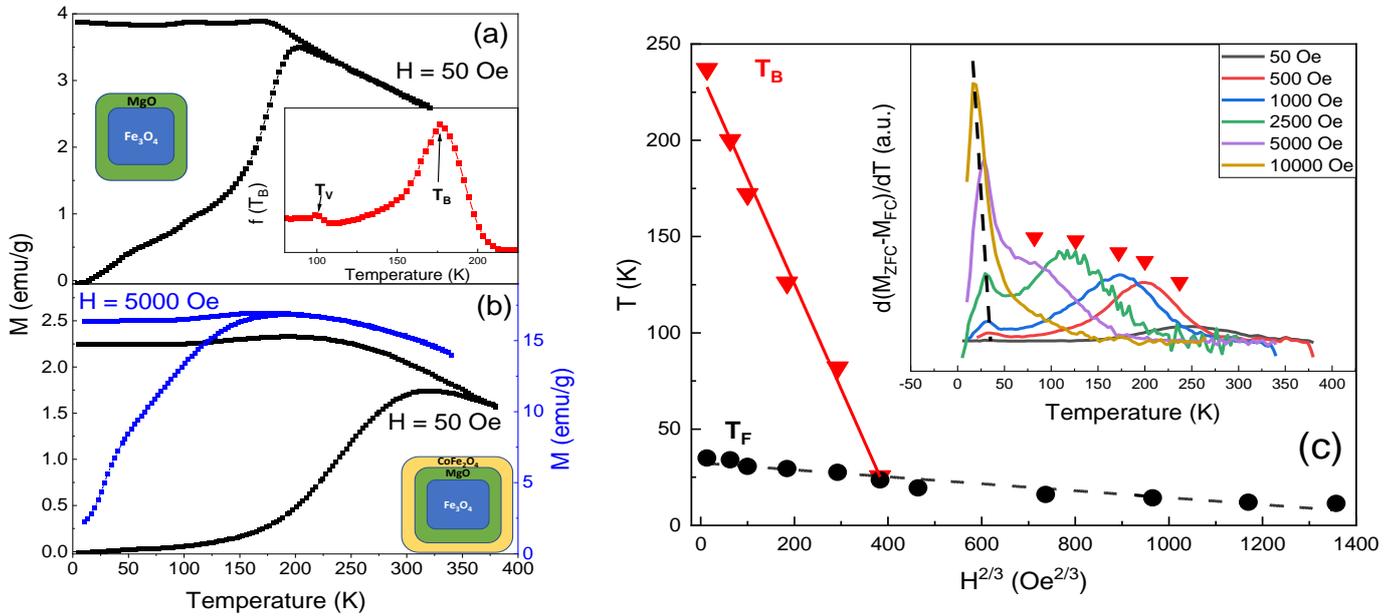

**Figure 7:** ZFC and FC temperature dependence of the magnetization curves of the core/shell (a) and core/shell/shell (b) nanoparticles, measured with H=50 Oe (black dots) and H=5000 Oe (blue dots). The inset shows $f(T_B) = \frac{1}{T}\frac{d(M_{ZFC}(T)-M_{FC}(T))}{dT}$ curve (red dots), where the maximum corresponds to the most probabe blocking temperature, $T_B$. (c) Field dependence of $T_B$ and the freezing temperature $T_F$ of the $Fe_3O_4/MgO/CoFe_2O_4$ nanoparticles systems. The inset shows the $\frac{d(M_{ZFC}(T)-M_{FC}(T))}{dT}$ used to determine $T_F$ (dash line), also the maximum associated to $T_B$ is signaled (red triangles).

Figure 7-(a) and (b) present the ZFC-FC magnetization curves of the $Fe_3O_4/MgO$ and the $Fe_3O_4/MgO/CoFe_2O_4$ nanoparticles systems, respectively. The magnetization of the $Fe_3O_4/MgO$ NPs shows a change from reversible to irreversible behaviour in agreement with the change from superparamagnetic to blocked regime. From this measurement the distribution of blocked temperature $f(T_B) = \frac{1}{T}\frac{d(M_{ZFC}(T)-M_{FC}(T))}{dT}$ can be calculated, where the most probable blocking temperature is obtained from the maximum of the distribution <$T_B$> = 177 K. From this figure it is also notice a kink at $T_V$= 101(2) K associated with the $Fe_3O_4$ Verwey transition, present at lower temperature compared to the bulk $T_V$ ~120 K,[70] due to size effects and deviation from stoichiometry.[71–74] When the $CoFe_2O_4$ is grown over the $Fe_3O_4$/MgO the irreversibility in the ZFC-FC magnetization is shifted to higher temperature signaling an increase of the energy barrier of the system in agreement with the magnetic coupling of both phases. From the $f(T_B)$ curves calculated from the magnetization of the $Fe_3O_4$/MgO/$CoFe_2O_4$ NPs measured with H=50 Oe (Fig. 7-(b)) the most probable blocking temperature is obtained resulting <$T_B$> =237 K. Notice that this value is in agreement with the one obtained from the fitting of the temperature dependence of $H_C$, shown in Fig. 6-(c). Also the non-monotonous behaviour of the core/shell/shell FC magnetization curve is noteworthy, an anomaly that is more evident when the measurement is collected applying larger magnetic fields. The inset of Fig. 7-(c) shows the $\frac{d(M_{ZFC}(T)-M_{FC}(T))}{dT}$ curves for different measuring applied field where, beside the broad peak associated to the distribution of blocking temperature, a narrower peak centered at $T_F$ with lower field dependence is clearly identified. Figure 7-(c) shows that both $T_B$ and $T_F$ have a linear dependence with $H^{2/3}$. This field dependence is consistent with the Almeida-Tholousse (AT) line $H^{2/3} \propto (1 - T/T_F)$ for spin glass transition [63,75,76] but also with the field dependence of the blocking temperature as predicted by Brown[77] and Dormann et al.[78] However, the emergence of exchange bias effect suggests the formation of spin glass state at low temperature as observed in several NPs systems.[63,65,68,79–84] The onsets of the freezing process can be identified by the low temperature increasing of the derivative curve at T~ 70 K in agreement with the appearance of exchange bias field, and the freezing temperature of the system can be obtained from the extrapolation to zero field of the AT-line resulting $T_F \sim 32$ K. To further explore the magnetic dynamics of the NPs we performed ac susceptibility measurements as a function of the temperature under different ac excitation frequencies.

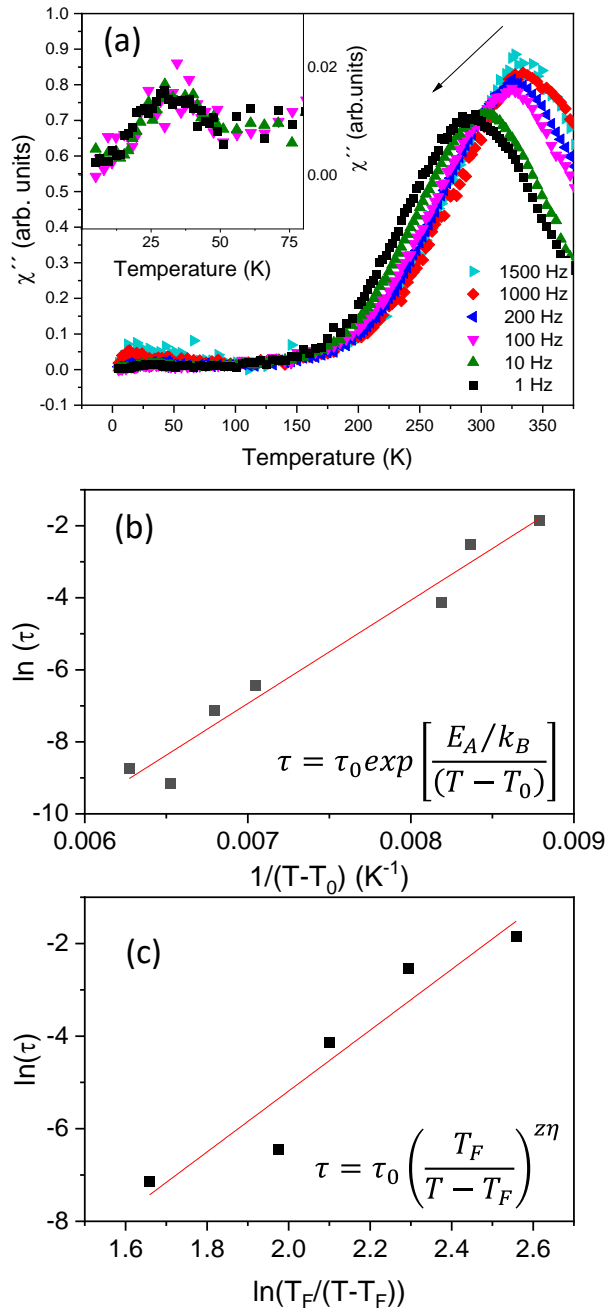

**Figure 8(a):** Imaginary component ($\chi''$) of the ac susceptibility measured ($H_{ac}=4$ Oe; 1 Hz$\leq$ f$\leq$ 1.5 kHz) for the $Fe_3O_4/MgO/CoFe_2O_4$ system. The inset shows a detail of the low

temperature region. (b) Frequency dependence of the high-temperature maximum and the fitting curve with the Vogel-Fulcher law. (c) Frequency dependence of the low-temperature maximum and the corresponding fit with a power law.

From the real ($\chi'$) and imaginary ($\chi''$) components of the ac magnetic susceptibility two different maxima in the data were observed at rather different temperatures. These 'peaks' are more clearly defined in the $\chi''(T)$ curve (see Figure 8-(a)), with a high temperature peak that shifts with increasing frequencies from T~293 K to T~ 333 K, and a low temperature peak that shifts from T~33.4 K to T~37 K in the same frequency range. We used the relative shift of temperature $T_m$ of the maximum in $\chi''(T)$ curves, per frequency decade $\varphi = \frac{\Delta T_m}{T_m \Delta \log(\omega)}$ as an indicator of the magnetic moment dynamic, obtaining $\varphi = 0.042$ and $0.032$ for the high- and low-maxima, respectively. These values show smaller frequency dependence than the expected for thermally activated superparamagnetic blocking mechanism in non-interacting NPs systems, and closer to values observed for spin glasses or strongly interacting single-domain magnetic NPs.[65,85,86] Consistently with these findings, the fitting of the high-temperature maxima using the Arrhenius law $\tau = \tau_0 \exp(E_A/k_B T)$, where $\tau = 1/2\pi f$ and $\tau_0$ is the characteristic relaxation time of the system, returns unreasonable physical results, evidencing the presence of interactions that affect the relaxation process. Therefore we used a phenomenological Vogel-Fulcher law $\tau = \tau_0 exp\left[\frac{E_A/k_B}{(T-T_0)}\right]$ to fit the frequency dependence of the high temperature peak, where $T_0$ account the interactions presents. From the fitting of the experimental results, shown in Fig. 8-(b), the following parameters were obtained: $\tau_0$ =2x10$^{-12}$ s, $E_A$=2869(300) K, and $T_0$=180 (20) K. This values are consistent with a thermally activated process of interacting magnetic moments, where the interactions are attributed to intraparticle effects due to the complex internal magnetic structure of the core/shell/shell nanoparticles. Instead, the low temperature peak is not consistent with a thermally activated process, and different model that account its slow dynamic should be applied. The critically slowing down relaxation time is usually model by a Power law $\tau = \tau_0 \left(\frac{T_F}{T-T_F}\right)^{z\eta}$, where $T_F$ is the static freezing temperature, and $zv$ is the dynamic exponent. Figure 8-(c) shows the frequency dependence of the low temperature maximum and the corresponding fit with a Power Law, from where $T_F$ = 31(6) K, $zv$ = 7(1), and $\tau_0$ = 1.5×10$^{-8}$ s were obtained. The parameters obtained from the fitting are in agreement with the reported for spin-glass, surface spin-glass like nanoparticles behaviour[61,82,87] and super spin-glass,[61,88] supporting the collective freezing model of the surface spins at low temperature.

The above results show that the onion-like nanoparticles present an enhancement of their coercivity $H_C$ compared with the single magnetic phase $Fe_3O_4$/MgO NPs, and the magnetization displayed a field dependence consistent with a single reversion that reflects the strong coupling between the $Fe_3O_4$ core with $CoFe_2O_4$ shell, notwithstanding the presence of the MgO spacer. This coupling could be due to surface irregularities, such as the FiM bridges observed by TEM where both FiM phases could be coupled by exchange interactions and, consequently, invert its magnetization together. Therefore, the magnetic moment of the onion-like nanoparticles behave superparamagnetically at room temperature and change to a blocked regime at $T_B$=237 K and below. Moreover, the doping of the cobalt ferrite with magnesium ions induces magnetic disorder evidenced by the reduction of the magnetic anisotropy of this phase and also by the presence of surface spins disorder that froze into a static and randomly oriented configuration at $T_F$=32 K. At the onsets of this spin glass transition, defined from the increases of the magnetization curve, emerge the exchange bias effect, manifested by the shifting of the FC magnetization loop and the enhancement of the coercivity field. The surface spin glass state is effective to pin the spins of the ferrimagnetic layer inducing large values of exchange bias field at low temperature as $H_{EB}$=2850 Oe at 5 K.

## CONCLUSIONS

We successfully fabricated three-layer $Fe_3O_4/MgO/CoFe_2O_4$ core/shell/shell magnetic nanoparticles using a thermal decomposition method, with an onion-like architecture confirmed by TEM and STEM microscopy. The resulting structure is formed by a magnetite core of 22 nm, encapsulated with an inner shell of MgO having ~1 nm thickness, and an outer cobalt ferrite shell of ~2.5 nm. We have identified the presence of FiM bridges through the diamagnetic MgO phase by EDS and EELS analysis, as well as partial Mg diffusion into the $CoFe_2O_4$ layer resulting in partial Mg-doping of the outer layer. The magnetic characterization showed that the magnetic moments of the $Fe_3O_4/MgO/CoFe_2O_4$ system fluctuate with a superparamagnetic regime in the time window of dc measurement (~100 s), changing to blocking regime at $T_B$=237K, higher than the $T_B$=177 K of the $Fe_3O_4/MgO$ nanoparticles, showing an effective magnetic coupling between the two magnetic phases through the MgO layer. An enhancement of the coercivity field is found when the third shell is grown due to the coupling between both FiM shells. The coupling is ascribed to the exchange interaction that is established through the MgO separator due to the presence of FiM bridges. At low temperature the disorder outer surface spins freeze in a spin glass state, that effectively pin the magnetic ions of the doped cobalt ferrite and, as a consequence, the system evidence exchange bias effects, which is manifested by the shifting and enhancement of the hysteresis cycle. The present results show the potential of the synthesis method for the design of new multiphase magnetic nanostructures in a single nanoparticle, and also it highlights the relevance of the structural, compositional, and interface details to the resulting magnetic phenomena at the level of individual particles.

## ASSOCIATED CONTENT

**Supporting Information**.

Additional data related to EELS spectra and EELS-SI mapping.

AUTHOR INFORMATION

**Corresponding Author**

*e-mail: winkler@cab.cnea.gov.ar

**Author Contributions**

The manuscript was written through contributions of all authors. All authors have given approval to the final version of the manuscript.

**ACKNOWLEDGMENT**

The authors acknowledge financial support of Argentinian governmental agency ANPCyT through Grant No. PICT-2019-02059, PICT-2018-02565 and UNCuyo for support through Grants No. 06/C604 and 06/C605. The authors gratefully acknowledge the EU Commission for financial support through MSCA-RISE projects #734187 (SPICOLOST) and #101007629 (NESTOR). SH was supported by the German Research Foundation (DFG project He 7675/1-1). G.F.G. thanks Spanish State Agency AEI for financial support through project PID2019-106947RB-C21. R.A. gratefully acknowledge the support from the Spanish MICINN (PID2019-104739GB-100/AEI/10.13039/501100011033), Government of Aragon (projects DGA E13-20R and from the European Union H2020 program "ESTEEM3" (Grant number 823717).

**Table of Contents Entry**

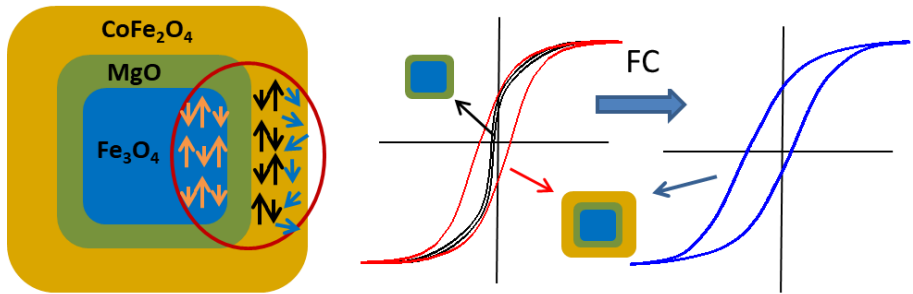

Nanoparticles with onion-like architecture offer a unique opportunity to modulate the coupling between magnetic phases by introducing spacers in the same structure. $Fe_3O_4/MgO/CoFe_2O_4$ shows enhanced coercivity due to the coupling between the FiM phases and exchange bias field originates from the freezing of the surface spins below the freezing temperature.